\documentclass[12pt]{iopart}
\usepackage{graphicx}

\begin{document}

\title[Entanglement in open quantum dynamics]{Continuous variable entanglement\\
in open quantum dynamics}

\author{Aurelian Isar}

\address{National Institute of Physics and Nuclear Engineering,
P.O.Box MG-6, Bucharest-Magurele, Romania}
\ead{isar@theory.nipne.ro}
\begin{abstract}
In the framework of the theory of open systems based on completely
positive quantum dynamical semigroups, we give a description of the dynamics of entanglement for a system consisting of two uncoupled harmonic oscillators interacting with a thermal environment. Using Peres-Simon necessary and sufficient criterion for separability of two-mode Gaussian states, we describe the evolution of entanglement in terms
of the covariance matrix for a Gaussian input state. For some values of the temperature of environment, the state keeps for all times its initial type:
separable or entangled. In other cases, entanglement generation, entanglement sudden death or a repeated collapse and revival of entanglement take place. We analyze also the time evolution of the logarithmic negativity, which characterizes the degree of entanglement of the quantum state.
\end{abstract}

%Uncomment for PACS numbers title message
% Keywords required only for MST, PB, PMB, PM, JOA, JOB?
%\vspace{2pc}
%\noindent{\it Keywords}: Article preparation, IOP journals
% Uncomment for Submitted to journal title message
%\submitto{\JPA}
% Comment out if separate title page not required
%\maketitle

\section{Introduction}

In recent years there is an increasing interest in using continuous variable entangled states in applications of quantum information processing and communication \cite{bra1}. A full characterization of the nonclassical properties of entangled states of continuous variable systems exists, at present, only for the class of Gaussian states. In this special case there exist necessary and sufficient criteria of entanglement \cite{sim,dua} and quantitative entanglement measures \cite{vid,gie}. The quantum information processing tasks are difficult to implement, due to the fact that any realistic quantum system is not isolated and it always interacts with its environment. Quantum coherence and entanglement of quantum systems are inevitably influenced during their interaction with the external environment. As a result of the irreversible and uncontrollable phenomenon of quantum decoherence, the purity and entanglement of quantum states are in most cases degraded. Therefore in order to describe realistically quantum information processes it is necessary to take decoherence and dissipation into consideration. Decoherence and dynamics of quantum entanglement in continuous variable open systems have been intensively studied in the last years \cite{oli,pra,ser1,ben1,avd,man,aphysa,aeur}.

When two systems are immersed in an environment, then, in addition to and at the same time with the quantum decoherence phenomenon, the environment can also generate a quantum entanglement of the two systems and therefore an additional mechanism to correlate them \cite{ben1,vvd1,ben2}.  Paz and Roncaglia \cite{paz1,paz2} also analyzed the entanglement properties of two oscillators in a common environment by using the exact master equation for quantum Brownian motion and showed that the entanglement can undergo three qualitatively different dynamical phases: sudden death, sudden death and revival, and no sudden death of entanglement.

In this paper we study, in the framework of the theory of open systems based on completely positive quantum dynamical semigroups, the dynamics of the continuous variable entanglement of two identical harmonic oscillators coupled to a common thermal environment.  We are interested in discussing the correlation effect of the environment, therefore we assume that the two oscillators are uncoupled, i.e. they do not interact directly. The initial state of the subsystem is taken of Gaussian form and the evolution under the quantum dynamical semigroup assures the preservation in time of the Gaussian form of the state. In section 2 we write the Markovian master equation in the Heisenberg representation for two uncoupled harmonic oscillators interacting with a general environment and the evolution equation for the covariance matrix of the considered subsystem. By using the Peres-Simon criterion for separability of two-mode Gaussian states \cite{sim,per}, we investigate in section 3 the dynamics of entanglement for this system. We show that for certain values of the environment temperature, the state keeps for all times its initial type: separable or entangled. For other values of the temperature, entanglement generation, entanglement sudden death or a repeated collapse and revival of entanglement take place. We analyze also the time evolution of the logarithmic negativity, which characterizes the degree of entanglement of the quantum state. A summary is given in section 4.

\section{Equations of motion for two harmonic oscillators}

We study the dynamics of the subsystem composed of two identical non-interacting oscillators in weak interaction with a thermal environment. In the axiomatic formalism based on completely positive quantum dynamical semigroups, the irreversible time evolution of an open system is described by the following general quantum Markovian master equation for an operator $A$ in the Heisenberg representation ($\dagger$ denotes Hermitian conjugation) \cite{lin,rev}:
\begin{eqnarray}{dA\over dt}={\rmi\over \hbar}[H,A]+{1\over
2\hbar}\sum_j(V_j^{\dagger}[A,
V_j]+[V_j^{\dagger},A]V_j).\label{masteq}\end{eqnarray}
Here, $H$ denotes the Hamiltonian of the open system and the operators $V_j, V_j^\dagger,$ defined on the Hilbert space of $H,$ represent the interaction of the open system with the environment.

We are interested in the set of Gaussian states, therefore we introduce such quantum dynamical semigroups that preserve this set during time evolution of the system. Consequently $H$ is taken to be a polynomial of second degree in the coordinates $x,y$ and momenta $p_x,p_y$ of the oscillators and $V_j,V_j^{\dagger}$ are taken polynomials of first degree in these canonical observables. Then in the linear space spanned by coordinates and momenta there exist only four linearly independent operators $V_{j=1,2,3,4}$ \cite{san}: \begin{eqnarray}
V_j=a_{xj}p_x+a_{yj}p_y+b_{xj}x+b_{yj}y,\end{eqnarray} where
$a_{xj},a_{yj},b_{xj},b_{yj}$ are complex coefficients. The Hamiltonian $H$ of the two uncoupled identical harmonic oscillators of mass $m$ and frequency $\omega$ is given by \begin{eqnarray}
H=\frac{1}{2m}(p_x^2+p_y^2)+{m\omega^2\over
2}(x^2+y^2).\end{eqnarray}

The fact that the evolution is given by a dynamical semigroup implies the positivity of the matrix formed by the scalar products of the four vectors
${\bi a}_x, {\bi b}_x,{\bi a}_y, {\bi b}_y$ whose entries are the components $a_{xj},b_{xj},a_{yj},b_{yj},$ respectively. We take this matrix of the following form, where all diffusion coefficients $D_{xx}, D_{xp_x},$... and dissipation constant $\lambda$ are real quantities (we put from now on $\hbar=1$):
\begin{eqnarray} \left(\begin{matrix}{D_{xx}&- D_{xp_x} -\rmi \frac{\lambda}{2}&D_{xy}& -
D_{xp_y} \cr - D_{xp_x} +\rmi \frac{\lambda}{2}&D_{p_x p_x}&-
D_{yp_x}&D_{p_x p_y} \cr D_{xy}&- D_{y p_x}&D_{yy}&- D_{y p_y}
- \rmi \frac{\lambda}{2} \cr - D_{xp_y} &D_{p_x p_y}&- D_{yp_y} + \rmi
\frac{\lambda}{2}&D_{p_y p_y}}\end{matrix}\right).\label{coef} \end{eqnarray}
It follows that the principal minors of this matrix are positive or zero. From
the Cauchy-Schwarz inequality the following relations hold for the coefficients defined in  (\ref{coef}): \begin{eqnarray}
D_{xx}D_{p_xp_x}-D^2_{xp_x}\ge\frac{\lambda^2}{4},~~~D_{yy}D_{p_yp_y}-D^2_{yp_y}\ge\frac{\lambda^2}{4},\nonumber\\
D_{xx}D_{yy}-D^2_{xy}\ge0,~~~D_{p_xp_x}D_{p_yp_y}-D^2_{p_xp_y}\ge 0,\nonumber\\
D_{xx}D_{p_yp_y}-D^2_{xp_y}\ge 0,~~~D_{yy}D_{p_xp_x}-D^2_{yp_x}\ge 0.
\label{coefineq}\end{eqnarray}

We introduce the following $4\times 4$ bimodal covariance matrix:
\begin{eqnarray}\sigma(t)=\left(\begin{matrix}{\sigma_{xx}(t)&\sigma_{xp_x}(t) &\sigma_{xy}(t)&
\sigma_{xp_y}(t)\cr \sigma_{xp_x}(t)&\sigma_{p_xp_x}(t)&\sigma_{yp_x}(t)
&\sigma_{p_xp_y}(t)\cr \sigma_{xy}(t)&\sigma_{yp_x}(t)&\sigma_{yy}(t)
&\sigma_{yp_y}(t)\cr \sigma_{xp_y}(t)&\sigma_{p_xp_y}(t)&\sigma_{yp_y}(t)
&\sigma_{p_yp_y}(t)}\end{matrix}\right).\label{covar} \end{eqnarray}
The problem of solving the master equation for the operators in Heisenberg representation can be transformed into a problem of solving first-order in time, coupled linear differential equations for the covariance matrix elements. Namely, from  (\ref{masteq}) we obtain the following system of equations for the quantum correlations of the canonical observables, written in matrix form \cite{san} ($\rm T$ denotes a transposed matrix):
\begin{eqnarray}{d \sigma(t)\over
dt} = Y \sigma(t) + \sigma(t) Y^{\rm T}+2 D,\label{vareq}\end{eqnarray} where
\begin{eqnarray} Y=\left(\begin{matrix} {-\lambda&1/m&0 &0\cr -m\omega^2&-\lambda&0&
0\cr 0&0&-\lambda&1/m \cr 0&0&-m\omega^2&-\lambda}\end{matrix}\right),\end{eqnarray}
\begin{eqnarray}D=\left(\begin{matrix}
{D_{xx}& D_{xp_x} &D_{xy}& D_{xp_y} \cr D_{xp_x}&D_{p_x p_x}&
D_{yp_x}&D_{p_x p_y} \cr D_{xy}& D_{y p_x}&D_{yy}& D_{y p_y}
\cr D_{xp_y} &D_{p_x p_y}& D_{yp_y} &D_{p_y p_y}} \end{matrix}\right).\end{eqnarray}
The time-dependent
solution of  (\ref{vareq}) is given by \cite{san}
\begin{eqnarray}\sigma(t)= M(t)[\sigma(0)-\sigma(\infty)] M^{\rm
T}(t)+\sigma(\infty),\label{covart}\end{eqnarray} where the matrix $M(t)=\exp(Yt)$ has to fulfill
the condition $\lim_{t\to\infty} M(t) = 0.$
In order that this limit exists, $Y$ must only have eigenvalues
with negative real parts. The values at infinity are obtained
from the equation \begin{eqnarray}
Y\sigma(\infty)+\sigma(\infty) Y^{\rm T}=-2 D.\label{covarinf}\end{eqnarray}

\section{Dynamics of two-mode continuous variable entanglement}

The characterization of the separability of continuous variable states using second-order moments of quadrature operators was given in Refs. \cite{sim,dua}. A Gaussian state is separable if and only if the partial transpose of its density matrix is non-negative [necessary and sufficient positive partial transpose (PPT) criterion]. A two-mode Gaussian state is entirely specified by its covariance matrix (\ref{covar}), which is a real, symmetric and positive matrix with the block structure
\begin{eqnarray}
\sigma(t)=\left(\begin{array}{cc}A&C\\
C^{\rm T}&B \end{array}\right),\label{cm}
\end{eqnarray}
where $A$, $B$ and $C$ are $2\times 2$ Hermitian matrices. $A$ and $B$ denote the symmetric covariance matrices for the individual one-mode states, while the matrix $C$ contains the cross-correlations between modes. Simon \cite{sim} derived a PPT criterion for bipartite Gaussian
continuous variable states: the necessary and sufficient criterion for separability is
$S(t)\ge 0,$ where \begin{eqnarray} S(t)\equiv\det A \det B+\left(\frac{1}{4} -|\det
C|\right)^2\nonumber\\- {\rm Tr}[AJCJBJC^{\rm T}J]- \frac{1}{4}(\det A+\det B)
\label{sim1}\end{eqnarray} and $J$ is the $2\times 2$ symplectic matrix
\begin{eqnarray}
J=\left(\begin{array}{cc}0&1\\
-1&0\end{array}\right).
\end{eqnarray}
Since the two oscillators are identical, it is natural to consider environments for which $D_{xx}=D_{yy},~ D_{xp_x}=D_{yp_y},~D_{p_xp_x}=D_{p_yp_y},~ D_{xp_y}=D_{yp_x}.$ Then both unimodal covariance matrices are equal, $A=B,$ and the entanglement matrix $C$ is symmetric.

\subsection{Time evolution of entanglement and logarithmic negativity}

In order to describe the dynamics of entanglement, we use the PPT criterion \cite{sim,per} according to which a state is entangled if and only if the operation of partial transposition does not preserve its positivity. Concretely, we have to analyze the time evolution of the Simon function $S(t)$ (\ref{sim1}). For a thermal environment characterized by the temperature $T,$ we consider such diffusion coefficients, for which \begin{eqnarray}m\omega D_{xx}=\frac{D_{p_xp_x}}{m\omega}=
\frac{\lambda}{2}\coth\frac{\omega}{2kT},~~~D_{xp_x}=0,\nonumber\\
m^2\omega^2D_{xy}=D_{p_xp_y}.\label{envcoe}\end{eqnarray} This corresponds to the case when the asymptotic state is a Gibbs state \cite{rev}.

For Gaussian states, the measures of entanglement of bipartite systems are based on some invariants constructed from the elements of the covariance matrix \cite{oli,avd,ovm}. In order to quantify the degrees of entanglement of the infinite-dimensional bipartite system states of the two oscillators it is suitable to use the logarithmic negativity.  For a Gaussian density operator, the logarithmic negativity is completely defined by the symplectic spectrum of the partial transpose of the covariance matrix. It is given by
$
L={\rm max}\{0,-\log_2 2\tilde\nu_-\},
$
where $\tilde\nu_-$ is the smallest of the two symplectic eigenvalues of the partial transpose $\tilde{{\sigma}}$ of the 2-mode covariance matrix $\sigma:$
\begin{eqnarray}2\tilde{\nu}_{\mp}^2 = \tilde{\Delta}\mp\sqrt{\tilde{\Delta}^2
-4\det\sigma}.
\end{eqnarray}
Here $ \tilde\Delta$ is the symplectic invariant (seralian), given by
$ \tilde\Delta=\det A+\det B-2\det C.$

In our model, the logarithmic negativity is calculated as \begin{eqnarray}L(t)=-\frac{1}{2}\log_2[4f(\sigma(t))], \end{eqnarray} where \begin{eqnarray}f(\sigma(t))=\frac{1}{2}(\det A +\det
B)-\det C\nonumber\\
-\left({\left[\frac{1}{2}(\det A+\det B)-\det
C\right]^2-\det\sigma(t)}\right)^{1/2}.\end{eqnarray}
It determines the strength of entanglement for $L(t)>0,$ and if $L(t)\le 0,$ then the state is
separable.

 In the following, we consider the behaviour of the logarithmic negativity for two types of the initial Gaussian state: 1) separable and 2) entangled.

1) In Figures 1 and 2 we represent the dependence of the logarithmic negativity $L(t)$ on time $t$ and temperature $T$ for a separable initial Gaussian state (unimodal squeezed state and respectively a mixed state). We notice that for relatively small values of the temperature $T,$ the initial separable state ($L(t)=0$) becomes entangled immediately or shortly after the initial moment of time $t=0$ (generation of entanglement, when the logarithmic negativity $L(t)$ becomes strictly positive). For relatively large values of $T,$ $L(t)$ keeps its initial zero value and the state remains separable for all times.

 Depending on the environment temperature, we notice three situations in the case of a generated entanglement: a) entanglement may persist forever, including the asymptotic final state; b) there exist repeated collapse and revival of entanglement; c) the entanglement is created only for a short time, then it disappears and the state becomes again separable. The entanglement of the two modes can be generated from an initial separable state during the interaction with the environment only for certain values of mixed diffusion coefficient $D_{xp_y}$ and dissipation constant $\lambda.$

\begin{figure}
{
\includegraphics{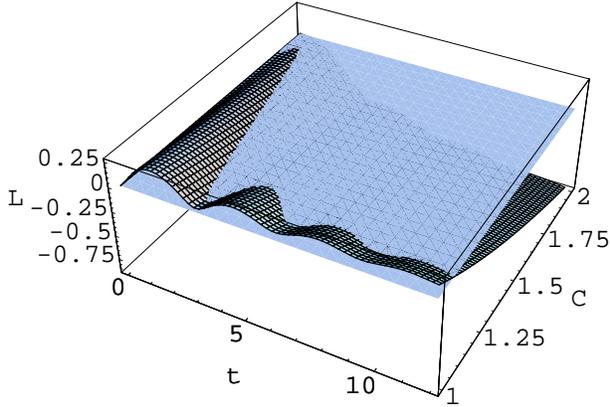}}
\caption{Logarithmic negativity $L$ versus time $t$
and environment temperature $T$ (via $C\equiv\coth\frac{\hbar\omega}{2kT}$) for $\lambda=0.1,$ $D_{xy}=0,$ $D_{xp_y}=0.049$ and separable initial uni-modal squeezed state with initial correlations $\sigma_{xx}(0)=3/4,~\sigma_{p_xp_x}(0)=1/3,~\sigma_{xp_x}(0)=\sigma_{xy}(0)=\sigma_{p_xp_y}(0)=~\sigma_{xp_y}(0)=0.$ We take $m=\omega=\hbar=1.$
}
\label{fig:1}
\end{figure}

\begin{figure}
{
\includegraphics{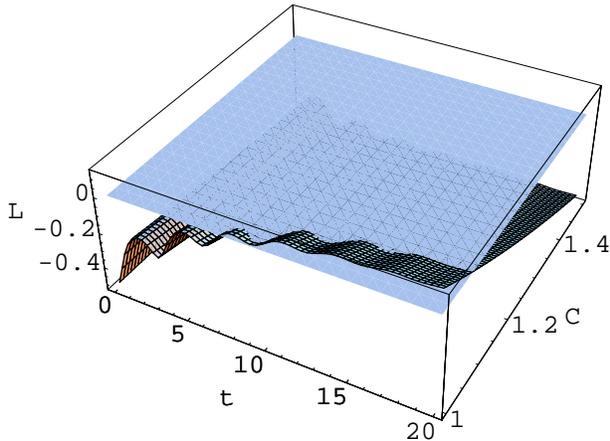}}
\caption{Same as in Figure 1, for a separable initial Gaussian mixed state with initial correlations $\sigma_{xx}(0)=1,~\sigma_{p_xp_x}(0)=1/2,~\sigma_{xp_x}(0)=\sigma_{xy}(0)=\sigma_{p_xp_y}(0)=~\sigma_{xp_y}(0)=0.$
}
\label{fig:2}
\end{figure}

2) The evolution of an entangled initial state is illustrated in Figures 3 and 4, where we represent the dependence of the logarithmic negativity $L(t)$ on time $t$ and temperature $T$ for an entangled initial Gaussian state (unimodal squeezed state and respectively a mixed state). We observe that for relatively small values of $T,$ the initial entangled state remains entangled almost for all times. For relatively large values of temperature $T,$ at some finite moment of time, $L(t)$ becomes zero and therefore the state becomes separable. This is the so-called phenomenon of entanglement sudden death. This phenomenon is in contrast to the loss of quantum coherence, which is usually gradual \cite{aphysa,arus}. Depending on the values of the temperature, it is also possible to have a repeated collapse and revival of the entanglement.

\begin{figure}
 {
\includegraphics{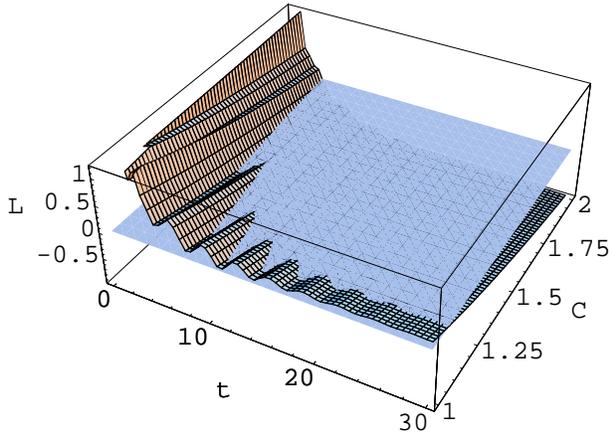}}
\caption{Same as in Figure 1, for an entangled initial uni-modal squeezed state with initial correlations $\sigma_{xx}(0)=3/4,~\sigma_{p_xp_x}(0)=1/3,~\sigma_{xp_x}(0)=0,~\sigma_{xy}(0)=1/2, ~\sigma_{p_xp_y}(0)=-1/2,~\sigma_{xp_y}(0)=0.$
}
\label{fig:3}
\end{figure}

\begin{figure}
 {
\includegraphics{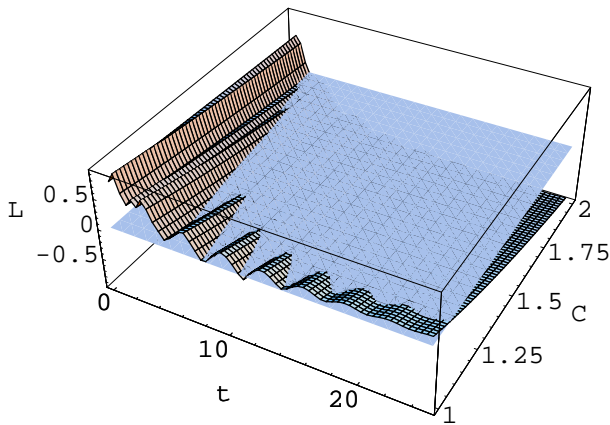}}
\caption{Same as in Figure 1, for an entangled initial Gaussian mixed state with initial correlations $\sigma_{xx}(0)=1,~\sigma_{p_xp_x}(0)=1/2,~\sigma_{xp_x}(0)=0,~\sigma_{xy}(0)=1/2, ~\sigma_{p_xp_y}(0)=-1/2,~\sigma_{xp_y}(0)=0.$
}
\label{fig:4}
\end{figure}

As expected, the logarithmic negativity has a behaviour similar to that one of the Simon function in what concerns the characteristics of the state of being separable or entangled \cite{arus,aijqi,ascri,aosid}.

\subsection{Asymptotic entanglement}

From (\ref{covarinf}) and (\ref{envcoe}) we obtain the following elements of the asymptotic matrices $A(\infty)=B(\infty):$
\begin{eqnarray} m\omega\sigma_{xx}(\infty)=\frac{\sigma_{p_xp_x}(\infty)}{m\omega}=\frac{1}{2}\coth\frac{\omega}{2kT}, ~~~\sigma_{xp_x}(\infty)=0
\label{varinf} \end{eqnarray}
and of the entanglement matrix $C(\infty):$
\begin{eqnarray}\sigma_{xy} (\infty) =
\frac{m^2(\lambda^2+\omega^2)D_{xy}+m\lambda
D_{xp_y}}{m^2\lambda(\lambda^2+\omega^2)},\end{eqnarray}
\begin{eqnarray}\sigma_{xp_y}(\infty)=
\sigma_{yp_x}(\infty)=\frac{\lambda
D_{xp_y}}{\lambda^2+\omega^2},\end{eqnarray}
\begin{eqnarray}\sigma_{p_xp_y} (\infty) =
\frac{m^2\omega^2(\lambda^2+\omega^2)D_{xy}-m\omega^2\lambda D_{xp_y}}{\lambda(\lambda^2+\omega^2)}.\end{eqnarray}
Then the Simon expression (\ref{sim1}) takes the following form in the limit of large times: \begin{eqnarray} S(\infty)=
\left(\frac{1}{4}(\coth^2\frac{\omega}{2kT}-1)-\frac{m^2\omega^2D_{xy}^2}{\lambda^2}+
\frac{D_{xp_y}^2}{\lambda^2+\omega^2}\right)^2\nonumber\\-\frac{D_{xp_y}^2}{\lambda^2+
\omega^2}\coth^2\frac{\omega}{2kT}.\label{sim2}\end{eqnarray}
For environments characterized by such coefficients that the expression $S(\infty)$ (\ref{sim2}) is strictly negative, the asymptotic final state is entangled. Just to give an example, without altering the general feature of the system, we consider the particular case $D_{xy}=0.$ Then, for a given temperature $T,$ we obtain that $S(\infty)<0,$ i.e. the asymptotic final state is entangled, for the following range of values of the mixed diffusion coefficient $D_{xp_y}$:
\begin{eqnarray}
\coth\frac{\omega}{2kT}-1<\frac{2D_{xp_y}}{\sqrt{\lambda^2
+\omega^2}}<\coth\frac{\omega}{2kT}+1.\label{insep}\end{eqnarray} We remind that, according to inequalities (\ref{coefineq}), the coefficients have to fulfill also the constraint \begin{eqnarray}\frac{\lambda}{2}\coth\frac{\omega}{2kT}\ge D_{xp_y}.\end{eqnarray} If the coefficients do not fulfil the double inequality (\ref{insep}), then $S(\infty)\ge 0$ and the asymptotic state of the considered system is separable.

The asymptotic logarithmic negativity has the form
\begin{eqnarray} L(\infty)=-\log_2\left[\left|\coth\frac{\omega}{2kT}-\frac{2D_{xp_y}}{\sqrt{\lambda^2
+\omega^2}}\right|\right].\end{eqnarray}
It depends only on the mixed diffusion coefficient, dissipation constant and temperature, and does not depend on the initial Gaussian state.

\section{Summary}

In the framework of the theory of open quantum systems based on completely positive quantum dynamical semigroups, we investigated the Markovian dynamics of the quantum entanglement for a subsystem composed of two noninteracting modes embedded in a thermal environment. By using the Peres-Simon necessary and sufficient criterion for separability of two-mode Gaussian states, we have described the evolution of entanglement in terms of the covariance matrix for Gaussian input states. For some values of diffusion and dissipation coefficients and of environment temperature, the state keeps for all times its initial type: separable or entangled. In other cases, entanglement generation or entanglement suppression (entanglement sudden death) take place or even one can notice repeated collapse and revival of entanglement. The dynamics of the quantum entanglement is sensitive to the initial states and the parameters characterizing the environment (diffusion and dissipation coefficients and temperature). We have also shown that, independent of the type of the initial state - separable or entangled, for certain values of temperature, the initial state evolves asymptotically to an equilibrium state which is entangled, while for other values of temperature the asymptotic state is separable. We described also the time evolution of the logarithmic negativity, which characterizes the degree of entanglement. For a given temperature, we determined the range of mixed diffusion coefficients for which the entanglement exists in the limit of long times.

\ack
The author acknowledges the financial support received within
the Project IDEI 497/2009 and Project PN 09 37 01 02/2009. I would also like to thank the referee for suggested recommendations to improve the quality of the paper.

\end{document}